# A Virucidal Face Mask Based on the Reverse-flow Reactor Concept for Thermal Inactivation of SARS-CoV-2

Samuel Faucher[1], Daniel James Lundberg[1], Xinyao Anna Liang[2], Xiaojia Cindy Jin[1], Rosalie Phillips[1], Dorsa Parviz[1], Jacopo Buongiorno[2], and Michael Strano[1]*

[1]Department of Chemical Engineering, Massachusetts Institute of Technology, Cambridge, MA, USA.

[2]Department of Nuclear Science and Engineering, Massachusetts Institute of Technology, Cambridge, MA, USA.

## Abstract

While facial coverings over the nose and mouth reduce the spread of the virus SARS-CoV-2 by filtration, masks capable of viral inactivation by heating could provide a complementary method to limit viral transmission. In this work, we introduce a new virucidal face mask concept based on a reverse-flow reactor driven by the oscillatory flow of human breath. The governing heat and mass transport equation are formulated and solved to analyze designs that evaluate both viral and $CO_2$ transport during inhalation and exhalation. Given limits imposed by the volume and frequency of human breath, the kinetics of SARS-CoV-2 thermal inactivation, and human safety and comfort, heated masks may inactivate SARS-CoV-2 in inflow and outflow to medical grade sterility. We detail one particular design, with a volume of 300 mL at 90 °C, that achieves a 3-log reduction in viral load with minimal viral impedance within the mask mesh, with partition coefficient around 2. This study is the first quantitative analysis of virucidal thermal inactivation within a protective face mask and addresses a pressing need for new approaches for personal protective equipment during a global pandemic.

**Topical Heading:** Transport Phenomena and Fluid Mechanics

## 1. Introduction

Face masks reduce the rate of person-to-person transmission of coronaviruses, influenza, and other respiratory viruses from breath and coughing.[1–3] During the COVID-19 pandemic, widespread adoption of effective masks has led to improved health outcomes around the world.[1,4] Face masks, however, are not uniformly effective in preventing person-to-person viral spread, and availability during the COVID-19 pandemic has often been limited.[3] An overwhelming majority of masks that have been employed during the 2020 pandemic, including N95 respirators, reduce viral transport by mechanical filtration at ambient temperature.[5] In contrast, there have been a dearth of versions designed around thermal viral inactivation and sterilization of air flow.

As an engineering problem, the cyclic reversal of air flow associated with human inhalation and exhalation enables a particular chemical reactor design: the reverse-flow reactor. A reverse-flow reactor[6–9] periodically reverses the direction of the convective feed through a one-dimensional reactor, typically a packed bed, to propagate a reactive zone over lengths that exceed the physical dimensions of the reactor. After the first patent filing by Frederick Cottrell in 1935,[10] reverse-flow reactors have been used industrially for nearly a half century and are well-studied.[6,8,11,12] This type of reactor offers several advantages when compared to unidirectional packed bed reactors. Depending on the frequency of flow direction switching, the reaction zone for an exothermic reaction can extend beyond the physical limits of the reactor in both directions, increasing conversion, reducing the need for heat exchange, limiting reactor fouling,[6] and improving stability to fluctuating inputs.[11] Coupling between heat and momentum transfer results in several periodic steady states.[12] While canonical reverse-flow reactors involve exothermic reactions, reverse-flow reactor designs can be applied to endothermic and mixed endothermic-exothermic systems as well.[12] In the case of a heated mask, the reaction (i.e. thermal inactivation of SARS-CoV-2) is not appreciably endothermic or exothermic, decoupling the temperature profile from the extent of reaction. In a strict sense, this eliminates some of the advantages of a reverse-flow reactor design. The reverse-flow concept in a virucidal mask, however, is imposed by the oscillations of human breath, not consciously selected by the engineer. Concepts taken from reverse-flow reactors, particularly the idea of distinct operating regimes delineated by bifurcation variables,[7] can be applied to a heated mask and yield insight to its design.

A heated mask to inactivate SARS-CoV-2 represents a new concept for personal protective equipment to address a critical and timely challenge. Thermal inactivation in a non-mask context has been widely discussed as a way to sterilize surfaces and objects to prevent transmission of the coronavirus.[13–15] Air circulation and heating of rooms and buildings has been proposed to reduce spread of SARS-CoV-2 in air as well.[16] With respect to masks, mechanical filtration is the main mechanism of pathogen impedance to date. N95 respirators and other mechanical filtration masks operate by interception, inertial impaction, and diffusion of particles, resulting in rejection of a majority of particles with sizes from the sub-micron scale to micron scale.[5] While N95 respirators are recommended for use in health care environments,[17] they are designed for single use and widespread adoption by non-healthcare workers can lead to shorages.[18] Reusable masks, such as cloth masks, offer substantial social benefit but may not offer adequate levels of protection from coronavirus in all circumstances,[19] while efforts to reuse masks (like N95 respirators) that are designed for single use may also reduce filtration efficiency.[20–22] Mask shortages, the generation of waste from single-use facemasks, and the spread of the COVID-19 pandemic to regions of the world with weaker healthcare infrastructure present an urgent need to reconsider designs and concepts for protective face masks.[3,23] In response, we propose a mask that blocks SARS-CoV-2 by thermal inactivation rather than mechanical filtration. Such masks have not been analyzed before in detail, making this the first quantitative analysis of the temperatures, volumes, and materials that could be used for thermal inactivation of a pathogen in a protective face mask.

In this work, we present a concept for a reusable facemask which contains a heated, porous mesh to thermally inactivate SARS-CoV-2. We formulate coupled mass and energy balances across a mask to create a design space that maps the operating temperature, mask volume, and air flow to viral load reduction and $CO_2$ accumulation. We analyze the wearable heated mask as a thermochemical reverse-flow reactor, and explore its design given three key model inputs: (1) oscillatory breath flow, (2) first-order thermal inactivation of SARS-CoV-2, and (3) a maximum allowable pressure drop. We propose a 300 mL mask that operates at a temperature of 90 °C, containing a copper mesh with an approximate mesh diameter of 0.1 mm. We show that this mask design can achieve a 3-log reduction in virus concentration with moderate viral impedance, in which virus particles travel one-third as fast as air inside the mask, or a 6-log reduction in virus concentration with higher viral impedance. By considering adsorption and desorption mechanisms and setting an overall pressure drop constraint, we determine that these partition coefficients can likely be achieved. These results show that a heated face mask is a promising, new design to reduce person-to-person spread of SARS-CoV-2, and will inform future prototyping and more detailed 3D modelling of pathogen thermal inactivation in heated face masks.

## 2. Problem Statement

A heated mask for thermal inactivation of SARS-CoV-2 contains a well-specified interior volume with a porous mesh, insulating outer layer, inlets to the bulk, and outlets to the nose and mouth, as shown in **Figure 1a-b**. This can be conceptualized as a 1D domain of length L, as shown in **Figure 1c-d**. The mask design is bounded by several constraints, including the volume and period of human breath, coronavirus inactivation kinetics, and a maximum pressure drop for safety and comfort. Within this design space, the optimal face mask is compact, achieves high viral inactivation, allows $CO_2$ exhalation, operates within a safe temperature range, and can be battery-powered for an extended time. These performance metrics, particularly the log reduction in viral concentration, depend on the mask volume, temperature, and a partition coefficient $K_p$. Five aspects of problem formulation – human breath, SARS-CoV-2 thermal inactivation, mass transfer, heat transfer, and viral impedance – are discussed below.

### 2.1 Human Breath

The oscillations of human breath impose a periodically reversing flow through the mask, in analogy with a reverse-flow reactor. The tidal volume, which is the volume of air displaced between inhalation and exhalation, is assumed to be 0.5 L,[24] while the period is assumed to be 5 seconds.[24] To ensure comfort, the pressure drop is held below 60 Pa L$^{-1}$ s$^{-1}$ at maximum flow, which represents the threshold for detection of inspiratory resistance.[25] The velocity within the mask, $U(t)$, is assumed to be sinusoidal:[26]

$$U(t) = U_{max} \cdot \sin\left(\frac{2\pi t}{\tau}\right) \quad [1]$$

Where $U_{max}$ is the maximum velocity and $\tau$ is the period. The assumed breath waveform is shown in **Figure 1e.** Positive velocities represent inhalation, and negative velocities represent exhalation.

### 2.2 Thermal Inactivation of SARS-CoV-2

The thermal inactivation of a virus, including SARS-CoV-2, follows well-characterized kinetics. It can be conceptualized as a first-order chemical reaction with a rate that varies with temperature as shown below:

$$r = -k(T)C \qquad [2]$$

where r is the reaction rate, $k$ is the first-order rate constant for thermal inactivation, and C is viral concentration. The rate of thermal inactivation of SARS-CoV-2 was fit following an Arrhenius relationship of the following form:

$$\ln(k) = -\frac{E_a}{RT} + \ln(A). \qquad [3]$$

Where $k$ is the first-order rate constant for thermal inactivation, $E_a$ is the activation energy for inactivation, $R$ the gas constant, $T$ the temperature, and $A$ the frequency factor. We fit experimental data obtained from Chin et al.[27] for SARS-CoV-2, as shown in **Figure 1f**, to find an activation energy of 132.6 kJ mol$^{-1}$ and a natural log of the frequency factor, $\ln(A)$, of 47.4. The correlation between the two fit parameters, $E_a$ and $\ln(A)$, follows a Meyer-Neldel rule that is suggestive of protein denaturation at high temperatures,[28] and is in agreement with the kinetics of thermal inactivation of a broad range of other coronaviruses, including the first Severe Acute Respiratory Syndrome coronavirus (SARS-CoV-1), Middle East Respiratory Syndrome (MERS-CoV), Transmissible Gastroenteritis Virus (TGEV), Mouse Hepatitis Virus (MHV), and Porcine Epidemic Diarrhea Virus (PEDV), as reported by Yap et al[28] and plotted in **Figure 1g**.

### 2.3 Mass Transfer

A 1D reaction-convection-diffusion model was used to analyze the concentration of virus and carbon dioxide inside a domain which extends through a mask from the mouth to bulk air. The governing equation for this system is as follows:[29]

$$\frac{\partial C(x,t)}{\partial t} = -U(t)\frac{\partial C(x,t)}{\partial x} + D_e \frac{\partial^2 C(x,t)}{\partial x^2} + r(C,x,t) \qquad [4]$$

where x is position, t is time, and $D_e$ is the dispersion coefficient.

The dispersion coefficient for the virus traveling within the mask is estimated as $5 \times 10^5$ m$^2$/s, taken from experimental evaluations of dispersion resulting from fluid flow through a mesh-screen packed column.[30] Virus concentration in exhaled breath is assumed to be zero, while virus concentration in the bulk is set to an arbitrary value $C_0$. During inhalation, Neumann boundary conditions are applied at the mouth and Danckwerts boundary conditions at the mask edge; these conditions are switched for exhalation.[31,32] Partial differential equations within the 1D mask domain were solved in MATLAB using function *pdepe* to obtain profiles of virus concentration, $CO_2$ concentration, and temperature during inhalation and exhalation.

### 2.4 Heat Transfer

The temperature distribution is governed by a similar equation to that of the concentration profiles. The temperature and concentration PDEs are not coupled: the temperature profile determines the concentration profiles, but the concentration profiles do not affect the temperature profile due to the negligible heat of reaction. The mask is assumed to contain a porous copper mesh which is heated by Joule heating, causing thermal inactivation of virus as well as slowing viral transport. Rapid thermal equilibrium between the mesh packing and air is assumed, while radiative heat transfer, work done by pressure changes, and viscous dissipation are ignored.[33] Validation of thermal equilibrium approximation is included in the supplementary section. Under these simplifications, the governing equation for the thermal distribution within a mask containing solid and fluid phases is as follows:

$$(\rho c)_m \frac{\partial T}{\partial t} + (\rho c_p)_f \varphi U_0 \frac{\partial T}{\partial x} = k_m \frac{\partial^2 T}{\partial x^2} + q - \frac{h_{eff} P_m (T - T_{amb})}{A_c} \quad [5]$$

$$(\rho c)_m = (1 - \varphi)(\rho c)_s + \varphi(\rho c_v)_f \quad [6]$$

$$k_m = (1 - \varphi) k_s + \varphi k_f \quad [7]$$

$$q = \frac{W}{A_c * L} \quad [8]$$

$$\frac{1}{h_{eff}} = \frac{1}{h_f} + \frac{L_{ins}}{k_{ins}} \quad [9]$$

The subscripts s and f refer to a solid phase (i.e. copper mesh) and a fluid phase (i.e. virus-laden air), respectively. $c$ is the specific heat of copper and is taken as 400 W mK$^{-1}$, $c_p$ is the specific heat at constant pressure of the fluid and is taken as 1020 J kg$^{-1}$ K$^{-1}$, $U_0$ is the superficial velocity of the fluid, $k_m$ is the effective thermal conductivity, $q$ is the heat production per unit volume, $W$ is the imposed electric power, $A_c$ is the mask flow area, $L$ is the length of the mask, $P_m$ is the perimeter, $T_{amb}$ is the ambient temperature taken as 20 °C, $h_{eff}$ is the effective heat transfer coefficient, $h_f$ is the free convection heat transfer coefficient for air on the outside surface of the mask and is taken as 2 W m$^{-2}$ K$^{-1}$, and $L_{ins}$ and $k_{ins}$ are the thickness and thermal conductivity of insulator material, which we assume for sake of comparison is neoprene and 0.3 cm thick. The porosity of the mesh, $\varphi$, is taken as a constant 0.9. During inhalation, the outflow Neumann boundary condition is applied at the mouth and the Danckwerts boundary condition at the mask edge.[31,32] During exhalation, the boundary conditions are flipped:

$$k_m \frac{dT(x=0, t_{inh})}{dx} A_f = \dot{m} C_p (T|_{x=0} - T_{amb}), \dot{m} > 0 \quad [10]$$

$$-k_m \frac{dT(x=a, t_{inh})}{dx} = 0 \quad [11]$$

$$k_m \frac{dT(x=a,t_{exh})}{dx} A_f = \dot{m} C_p (T|_{x=a} - T_{body}), \dot{m} < 0 \qquad [12]$$

$$-k_m \frac{dT(x=0,t_{exh})}{dx} = 0 \qquad [13]$$

Physically speaking, heat transferred at the outer surface of the mask preheats the cold air that enters the mask when inhaling (**Equation 10**) when exhaling, air that is breathed out also heats up due to the heat transfer at the mask-mouth interface (**Equation 11**). The temperature of the air entering the mask is taken to be 20 °C, and the temperature of the exhaled air is taken to be 37 °C.

### 2.5 Viral Impedance

Virus in the mask is subject to thermal inactivation, but it is also impeded by the porous mesh as in other masks that operate purely by filtration. Numerous mathematical models have been developed to predict the particle transport and retention in porous media, using either macroscopic or microscopic approaches.[34,35] Macroscopic methods use the particle transport equation in a continuous media:

$$\varphi \frac{\partial C}{\partial t} + \frac{\partial (UC)}{\partial x} = -\Lambda U C \qquad [14]$$

in which $\Lambda$ is the filtration coefficient, which is related to many parameters including pore structure, particle size distribution, and the particle-surface interactions that govern the particle adsorption and release on/from the surface. Theoretical calculation of $\Lambda$ is very difficult, so its value is usually determined experimentally.[36–38]

On the other hand, microscopic approaches investigate the particle retention at pore scale, using direct models such as CFD-DEM6 or, more recently, pore network modeling.[39–42] Pore network modelling starts with force balances on a single particle, accounting for hydrodynamic drag, body force, electrostatic, van der Waals, and inertial force and relating the particle velocity to fluid velocity:

$$U_p = U - \frac{1}{6\pi\mu R}(F_B + F_E + F_V) \quad [15]$$

in which $U_p$ is the particle velocity, $F_B$ is body force, $F_E$ is the electrostatic force and $F_V$ is the van der Waals force. Estimates for each force term can be determined from particle and surface potentials. In addition to advection, the Brownian diffusion of virus must also be included in its microscopic transport formulation. Additionally, particle retention in the porous mesh occurs via particle adsorption and binding to the mesh. Each of these steps may be explained by a combination of phenomena including gravitation sedimentation, Brownian motion, and surface forces. Eventually, these single-particle equations must be applied to an interconnected network of pores and throats.[43] In the absence of an experimentally measured filtration coefficient and other parameters for SARS-CoV-2 and to avoid the complicated pore network modeling, a simple model relating the virus velocity to bulk air velocity must be used. We introduce a simple model of viral impedance within the mask, where the virus is subject to slowing according to a partition coefficient $K_p$ as shown below:

$$U_{vir} = \frac{U}{1+K_p} \quad [16]$$

In this case, $U_{vir}$ is the effective virus velocity and $U$ is the bulk air velocity. Viral impedance occurs because virus particles adsorb to the copper mesh and desorb from the copper mesh in two first-order processes. The partition coefficient $K_p$ can be defined as the ratio of these two rates:

$$K_p = \frac{k_{ads}}{k_{des}} \quad [17]$$

where $k_{ads}$ is the virus adsorption rate constant, while $k_{des}$ is the virus desorption rate constant. Mask performance metrics, including viral inactivation and $CO_2$ concentration, can be calculated as a function of mask volume, mean temperature, and partition coefficient $K_p$.

## 3. Results and Discussion

### 3.1 Mask Design

**Equation 4** and **Equation 5**, which govern temperature, virus concentration, and $CO_2$ concentration within the mask as function of position and time, were solved in Matlab using function *pdepe*. Time-variant virus, $CO_2$, and temperature profiles for a 300 mL mask operating at a mean temperature of 90 °C are shown in **Figure 2**, with inhalation in **Figure 2a-c** and exhalation in **Figure 2d-f**. Boundary conditions, as described above, fix the virus and $CO_2$ concentrations at the mask-bulk air interface ($x = 0$) and mask-mouth interface ($x = L$) during inhalation and exhalation, in accordance with the coordinate system presented in **Figure 1c**. The $CO_2$ concentration of exhalation was fixed at the mask-mouth interface at 3.8%, while the $CO_2$ concentration in bulk air is essentially zero. During inhalation, in **Figure 2a,** virus is inhaled but does not reach the mask-mouth interface due to viral impedance and thermal inactivation. A large majority of non-inactivated virus is exhaled, as shown in **Figure 2d.** $CO_2$, by contrast, is transported faster and is not subject to thermal inactivation. Air with near-zero $CO_2$ concentration is inhaled in **Figure 2b**, while air with high $CO_2$ concentration is exhaled in **Figure 2e** and transported across the mask to the bulk. For this design, a mean temperature of 80 °C, as shown in **Figure 2c** and **Figure 2f**, is achieved with a power input of 18.89 W. Concentration and temperature profiles for candidate designs with different volumes and temperatures are similar to the profiles shown.

Power requirements, average inhaled $CO_2$ concentration, and log viral inactivation are shown in **Figure 3** for masks ranging from 0.1 L to 1 L in volume, operating temperatures ranging from 40 °C to 140 °C, and $K_p$ ranging from 0 to 10. Power requirements range from 3.79 W for a 0.1 L mask at 40 °C to 52.96 W for a 1 L mask at 140 °C. $CO_2$ concentration, as expected, does not vary with temperature, but varies strongly with mask volume. Given the poor mixing in the mask, the mask volume must be less than the human tidal volume (500 mL) in order to achieve a non-hazardous $CO_2$ concentration.[44]

Performance of heated masks in achieving specified log reductions in SARS-CoV-2 is presented in **Figure 4**. The required operating temperature to induce a specified viral load inactivation is plotted as a function of mask volume for a range of values of the partition coefficient $K_p$. Two thresholds for viral inactivation are shown: a 3-log, or thousand-fold, reduction in inhaled viral concentration, and a 6-log, or million-fold, reduction in inhaled viral concentration. In general, 6-log reduction is the standard for sterilization.[45] In all cases, the mask aspect ratio – defined as the ratio of mask length to the geometric mean of cross-sectional dimensions – was set to 3.

A 3-log viral reduction can be achieved in a compact mask (<0.5 L) at a reasonable operating temperature (<100 °C) if viral transport is impeded moderately in the mask ($K_p > 1$), as shown in **Figure 4a**. A 6-log viral reduction can be achieved within the same volume and temperature constraints with slightly higher virus impedance in the mask ($K_p > 2$), as shown in **Figure 4b**. Without viral impedance, where $K_p = 0$, temperatures exceeding 100 °C or volumes approaching 1 L are required to achieve 3-log viral reduction, while temperatures exceeding 100 °C and volumes exceeding 1 L are required to achieve 6-log viral reduction. The case of no viral impedance represents a worst-case scenario for mask function, as any reasonable mesh design would combine thermal inactivation with filtration and particle impedance to some degree.

Two regimes are apparent in **Figure 4**: At low temperatures, below 90 °C, viral reduction is due largely to viral impedance. Vertical lines in this range show that there is little temperature dependence, with large differences in required volume for different levels of $K_p$. By contrast, at high temperatures above 90 °C, viral reduction is due largely or entirely to thermal inactivation. The inflection and flattening of curves in this range show that the degree of viral reduction is substantially temperature dependent above 90 °C but decreasingly dependent on viral impedance, with only minor differences in required volume for different levels of $K_p$. At high temperatures, the virus is thermally inactivated so rapidly that viral reduction obtained for lower and higher values of $K_p$ become nearly identical. While this is extremely promising for viral inactivation, operating at temperatures above 100 °C presents issues for power requirements and human safety. In general, the presence of these two regions and the transition between them suggests a tradeoff between heated mask volume and heated mask temperature. A small, hot mask can achieve a similar level of viral inactivation as a larger, cooler mask. The secondary tradeoff, however, is in $CO_2$ levels and user comfort; a small mask allows for lower $CO_2$ levels during inhalation, while a large mask can be operated at a milder temperature.

The optimal mask design is located at the knee of the curves in **Figure 4a-b**. This volume and temperature will make maximum use of both thermal inactivation and traditional filtration to reduce viral transport and ensure safety for the mask wearer and those around them. The mask volume also determines the $CO_2$ concentration: a mask volume below the tidal volume of 500 mL is necessary to allow sufficient $CO_2$ transport for comfortable breathing. Based on these criteria, we choose a mask volume of 300 mL and an operating temperature of 90 °C. Renderings of a 300 mL mask, showing a porous copper mesh interior, neoprene insulation, two air inlets on the sides of the mask, and an outlet to the nose and mouth, are shown in **Figure 1a** and **Figure 1b**. Full details of the proposed design are compiled in Table S1.

### 3.2 Mask Performance Regimes

In analogy with reverse-flow chemical reactors,[9] the mask design space can be split into distinct regimes with qualitative differences in performance, as shown in **Figure 5**. These four regimes are characterized by two dimensionless groups: the dispersion number (Di) and the reduced volume ($\hat{V}$). The reduced volume $\hat{V}$ is the ratio of mask volume to breath tidal volume; in the case of a reverse-flow reactor, this is the ratio of the residence time to the switching time:

$$\hat{V} = \frac{V_{mask}}{V_T} \qquad [18]$$

The dispersion number captures dispersion of reactants in a reactor. A reactor with $Di \gg 1$ will behave as a stirred-tank reactor, while a reactor with $Di \ll 1$ will behave as a plug-flow reactor:

$$Di = \frac{D_e}{UL} \qquad [19]$$

In this case, the **$Di \ll 1$  $\hat{V} > 1$** limit, as shown in **Figure 5a**, has a large mask volume and is poorly mixed. While this means that the virus is not transported from the bulk to the mouth, it also does not allow transport of $CO_2$ out. We can call this the *plastic bag limit*, and it is not suitable for a protective facemask. There is a sharp bifurcation at $\hat{V} = 1,$ beyond which is the **$Di \ll 1$  $\hat{V} < 1$** regime, as shown in **Figure 5b**. In this regime, the mask volume is small but poorly mixed, leading to near-complete transport of $CO_2$ out of the mask at the end of exhalation. This *straw limit* is promising for a heated mask, provided that thermal inactivation of virus is suitably fast. In the *well-mixed limit*, with **$Di \gg 1,$** there is no strong qualitative difference at $\hat{V} = 1.$ These cases are shown in **Figure 5c** and **Figure 5d**. In both a small mask (**Figure 5c**) and large mask (**Figure 5d**), the high level of mixing means that there are virtually no spatial gradients in virus and $CO_2$ concentration, but there are strong temporal gradients during the breath cycle. In these cases, virus concentration is maximized after inhalation, while $CO_2$ concentration is maximized after exhalation.

Given the flow rate of human breath and dispersion coefficients of air in porous media, the dispersion number in the proposed mask designs is likely to be small regardless of the exact mask volume and operating temperature, so the well-mixed cases with $Di \gg 1$ are not immediately physically relevant. Nonetheless, it is important to compare the problem of heated mask design problem to these regimes of reverse-flow reactor transport. In a general sense, we can see that the ideal heated mask would impede virus transport but not $CO_2$ transport, combining filtration and thermal inactivation to act in the "*plastic bag*" limit toward virus particles but the "*straw*" limit toward exhaled $CO_2$.

### 3.3 Pressure Drop and Viral Impedance

The success of the proposed mask in inactivating SARS-CoV-2 relies on achieving viral impedance, as captured by $K_p$, without exceeding a maximum pressure drop. The structure and solidity of the mask mesh are critically important for both criteria. Specifically, the pressure drop constraint specifies a minimum mesh size for a given mask volume, while the mesh size dictates the expected viral impedance value, $K_p$. As shown in **Figure 6a-b,** our analysis suggests that in a 300 mL mask with a solidity of 0.1 and a copper mesh with diameter 0.1 mm will likely be able to achieve a viral impedance value of $K_p = 2$. This supports the conclusion that our proposed mask, with a size of 300 mL at an operating temperature of 90 °C, will be able to cause a 3-log reduction in viral concentration.

With respect to pressure drop, we set an upper bound of 60 Pa L$^{-1}$ s$^{-1}$ at maximum flow, a conservative lower threshold for noticeable inspiratory resistance.[46,47] We take the interior of the mask to be filled with a mesh composed of stacked screens of cross-woven fibers. The mesh is defined by two variables – the fiber diameter, $d_w$, and fiber spacing, $s_w$ – where the distance between stacked screens is taken to be equal to this fiber spacing, as shown in **Figure S2**. The pressure drop across a single layer of mesh screen under oscillatory flow is predicted from an empirical correlation:[48]

$$\Delta P = \frac{\rho U^2}{2} \frac{\alpha}{(1-\alpha)^2} (17 \cdot Re_d^{-1} + 0.55) \qquad [20]$$

where $\rho$ is the fluid density, $U_{max}$ the maximum superficial velocity during the oscillatory flow cycle, and $Re_d$ is the Reynolds number based on the wire diameter. The solidity of the screen, $\alpha$, is defined as the complement of the mesh porosity. For the specified geometry, the solidity of a single screen is equal to:

$$\alpha = \frac{d_w(2s_w+d_w)}{(s_w+d_w)^2} \qquad [21]$$

The average solidity of the stacked fiber screens is then given as:

$$\alpha' = \alpha \frac{d_w}{s_w+d_w} \qquad [22]$$

The pressure drop across more than a single mesh – in this case, across the porous interior of a mask—is taken to scale linearly with the number of mesh screens present, and thus scales linearly with length.[49] Given a mesh solidity of 0.1, the minimum wire diameter before the pressure drop constraint is violated for a mask design of some volume and cross-sectional area is shown in **Figure 6a**.

Given a wire diameter of 0.1 mm, it is likely that a partition coefficient of $K_p = 2$, which is required to achieve 3-log reduction in virus concentration, can be achieved. The maximum value of the partition coefficient, $K_p$, is shown as a function of wire diameter and cross-sectional area in **Figure 6b**. This predicted value depends on estimates for both the adsorption rate, $k_{ads}$, and the desorption rate, $k_{des}$. The adsorption rate can be calculated as shown below:

$$k_{ads} = \eta_f U \frac{4\alpha}{\pi d_w (1-\alpha)} \qquad [23]$$

where $\eta_f$ is a single-fiber efficiency. Captured particles are retained by Van der Waals forces,[50] but the mechanism by which they come into contact with the filter medium is dependent on filter and particle geometry. For coronavirus particles, which have a diameter of roughly 100 nm, the dominant mechanism of capture is through Brownian diffusion.[51] The governing parameter for this diffusive mechanism is the Péclet number, defined as the ratio of convective to diffusive transport rates:

$$Pé = \frac{U_0 d_w}{D} \qquad [24]$$

Where $d_w$ is the filter fiber diameter, and $D$ is the diffusion coefficient of the particle. The single fiber efficiency due to diffusion is obtained from an appropriate correlation such as that proposed by Wang et al.[52]:

$$\eta_f = 0.84 \cdot Pe^{-0.43} \qquad [25]$$

The diffusion coefficient for the particle is obtained from the Stokes-Einstein equation:

$$D = \frac{k_B T C_s}{3\pi\mu d_p} \quad\quad [26]$$

Where $k_B T$ is the product of the Boltzmann constant and temperature, $\mu$ is the fluid dynamic viscosity, $d_p$ the diameter of the particle, and $C_s$ the Cunningham slip correction factor:

$$C_s = 1 + K_n \left[1.207 + 0.44 \cdot exp\,exp\left(-\frac{0.78}{Kn}\right)\right] \quad\quad [27]$$

Where $K_n$ is the Knudsen number of the particle defined as:

$$K_n = \frac{2\lambda}{d_p} \quad\quad [28]$$

Where $\lambda$ is the mean free path of a gas molecule in the fluid. For air, and assuming ideality, the mean free path is obtained from the following correlation:

$$\lambda = RT/\sqrt{2}\pi d_{N_2} N_A P \quad\quad [29]$$

Where R is the gas constant, $d_{N_2}$ is the diameter of nitrogen, the dominant molecular species in air, $N_A$ is Avogadro's number, and $P$ the pressure.

The desorption rate, by contrast, is less well-known, so we assume a conservative value of $k_{des} = 1\,s^{-1}$ in our calculations. The desorption rate can likely be tuned by various mechanisms. One approach to tailor the desorption rate of captured particles includes making use of the natural oscillatory flow within the mask. Pulsed air flow can effectively cause particle desorption, and is used industrially to clean surfaces from particle debris.[53,54] Mechanical stimulus of the mesh could actively control particle desorption,[55] where optical[56] and pneumatic[57] actuation of these structures is also possible. Applied electric fields can also be used in the same way. By applying an electric field to a conductive mesh, an electrostatic desorption force can be induced.[58] This principle has been applied with success to biological filtration membranes with a continuously applied electrical field,[59] as well with pulsed electric fields to remove microbial surface contamination.[60]

Given these estimates for the adsorption and desorption rate, a partition coefficient $K_p$ of 2 is achievable for a 300 mL mask with a 0.1 mm mesh for a mask cross-sectional area of 50 cm², as shown in **Figure 6b.** From this analysis of adsorption and desorption rates and mechanisms, then, we see that the necessary viral impedance is likely achievable within the pressure drop and volume constraints of the design process, even though a cross-sectional area of 50 cm² fixes a different aspect ratio (AR=0.85) than the specified aspect ratio of 3. In general, there is a tradeoff between mask volume and achievable viral impedance. The attainable value of $K_p$ is a strong function of the cross-sectional area of the mask. A smaller cross-sectional area produces higher superficial velocities of flow, resulting in a larger value of the adsorption rate constant. However, these higher velocities also induce large pressure drop, and limit the maximum volume of mesh. Thus, at a given temperature, a smaller volume mask with a larger value of $K_p$ may be equally able to achieve a desired viral reduction as a larger mask with a lower value of $K_p$—where ultimately a smaller mask volume would be the overall preferable option due to lower required energy consumption and smaller profile.

### 3.4 Heating and Safety

Heating the mask requires continuous power input to maintain a set temperature. The incorporation of a power supply within any mask design requires considerations and tradeoff between the duration of power supply and the bulk and weight it adds to the design. **Figure 6c** shows the power supply required of a 300 mL mask over a range of set temperatures. This energy requirement is compared against the capacity of standard consumer batteries.

For a moderately heated mask of 90 °C, operation for an hour requires 20.7 W-hrs which can be supplied by two C batteries (130 grams total), six AA batteries (120 grams total), or twelve AAA batteries (108 grams total). Further, instead of single-use batteries, rechargeable batteries or battery pack would provide a lower cost and long-term way to power the mask. It is mentioned that the energy required to heat a mask at a given temperature scales nearly linearly with volume.

With respect to mask safety, continuous inhalation of and exposure to heated air and heating elements are matters of concern for the overall comfort and safety of any heated mask design. Appropriate insulation is necessary to limit the surface temperature of the mask—either on the face, or outwards—as well as minimize heat loss. Low-density but high thermally insulating neoprene is one material well-suited out of which the mask may be constructed. Ultra-insulating neoprene with thermal conductivity as low as 0.03 W m$^{-1}$ K$^{-1}$ has recently been constructed, by incorporating noble gases into neoprene foam[61]. Further, neoprene is heat-resistant up to temperatures of 135 °C,[62] making it an ideal material for use in the high-temperature regions of the mask.

Minimizing the temperature of inlet air which is exposed to the face and inhaled is also necessary to provide a safe and comfortable mask. Continual inhalation of air exceeding 125 °F (52 °C) or nearly has been shown to be painful for humans.[63] Accordingly, any mask design must incorporate a system to appropriately cool down air before inhalation.

The reduction in temperature of inhaled air can be accomplished through incorporation of low thermal diffusivity (the ratio of thermal conductivity to the product of density and specific heat capacity) materials at the inlet and outlet of the heated chamber of the mask. These regions of the mask containing these thermal masses would not be actively heated, but instead retain the thermal energy of the heated air—cooling it down before exiting the mask. This type of thermal design was explored as discussed in the Supporting Information. Notably, for a 300 mL mask heated to an average temperature of 90 °C, the electric power consumption was able to be reduced from 20.7 W to 15.8 W.

### 3.5 Further Improvements

There are mechanisms beyond thermal viral inactivation that may further improve the mask design. Coating the mesh with a layer of a metal such as copper or iron, which as ions have been shown widely to inactivate viruses and bacteria,[64,65] may improve mask performance. Anti-pathogenic polymeric[66] or non-polymeric[67] coatings may also increase the inactivation rate of adsorbed pathogens. Physical patterning, either on the microscale or nanoscale, could also lead to enhanced rate of inactivation,[68,69] as could the incorporation of UV light[70] with due caution about the hazards of ozone on human health.[71]

The long lifetime of the designed mask relies on its continued ability to inactivate and adsorb viral particles. The use of a small wire spacing within the mesh would minimize the potential of larger particles—such as dust or other debris that can be as large as 100 microns in diameter[72]—to cake or block individual pores on the filter.[73] The cyclic nature of breath, too, presents a natural mechanism of filter clearing. Exhalation, when the flow reverses direction within the mask, would backwash the filter medium, potentially releasing and expelling any particle contaminants present.[74] In general, care must be taken to design system which is able to retain, inactivate, and then emit that inactivated pathogen of interest without irreversibly capturing other contaminants. In the case where periodic replacement of the mesh material within the mask is necessary, a cheaper and less complex mesh material would be favored.

### 4. Conclusions

We have computationally analyzed the effectiveness of a heated face mask for thermal inactivation of the novel coronavirus, SARS-CoV-2. This constitutes the first quantitative analysis of face mask air purification by thermal inactivation of a pathogen. Relevant design choices for a heated mask include the mask volume, operating temperature, aspect ratio, and the diameter and spacing of fibers within the mask. Parameters which cannot be changed, or which follow from the design choices above, include the tidal volume and frequency of human breath, kinetics of coronavirus inactivation, human comfort and safety limits, power requirements, and the degree to which particles are impeded within the mask, as modelled by the partition coefficient $K_p$.

We have introduced a framework to consider these factors in mask design optimization, with an optimal mask volume of 300 mL at an operating temperature of 90 °C. This mask can achieve a 3-log reduction in virus concentration with $K_p = 2$, or 6-log reduction in virus concentration with $K_p = 5$. These values of the partition coefficient can likely be achieved without excessive pressure drop in a copper mesh with a wire diameter of roughly 0.6 mm. In general, we note a tradeoff between mask temperature and volume; the dependence of maximum mask volume on mesh material and spacing by the pressure drop constraint; and the presence of different regimes in heated mask design, which we call the plastic bag, straw, and well-mixed limits, on the basis of different dimensionless groups by analogy with reverse-flow chemical reactors. Given physical and safety limits, heated masks are promising options to protect against transmission of COVID-19 if coupled with thermoelectric cooling and thermally insulating liner materials. Future work, including experimental study with heated mask prototypes or more extensive 3D computational fluid dynamics simulations, will shed more light on the exact volume, temperature, and mesh size that are most promising for heated mask design, as well as the materials and battery systems that are most capable of achieving target viral inactivation.

# Notation

| | |
|---|---|
| $A$ | Frequency factor [1/s] |
| $A_c$ | Mask flow area [m²] |
| $c$ | Specific heat of copper [W m/K] |
| $c_p$ | Specific heat at constant pressure of the fluid [J/(kg K)] |
| $C$ | Virus concentration [mol/L] |
| $C_0$ | Virus concentration in the bulk air [mol/L] |
| $C_s$ | Cunningham slip correction factor [dimensionless] |
| $d_{N_2}$ | Diameter of nitrogen [m] |
| $d_p$ | Diameter of virus particle [m] |
| $d_w$ | Fiber diameter [m] |
| $D$ | Diffusion coefficient of the virus particle [m²/s] |
| $D_e$ | Dispersion coefficient [m²/s] |
| $Di$ | Dispersion number |
| $E_a$ | Activation energy for virus thermal inactivation [kJ/mol] |
| $F_B$ | Body force [N] |
| $F_E$ | Electrostatic force [N] |
| $F_V$ | Van der Waals force [N] |
| $h_{eff}$ | Effective heat transfer coefficient [W/m²/K] |
| $h_f$ | Free convective heat transfer coefficient for air on the outside surface of the mask [W/(m² K)] |
| $k$ | First-order rate constant for virus thermal inactivation [1/s] |
| $k_{ads}$ | Virus adsorption rate constant [1/s] |
| $k_B$ | Boltzmann constant [J/K] |
| $k_{des}$ | Virus desorption rate constant [1/s] |
| $k_{ins}$ | Thermal conductivity of insulating material (neoprene) [W/(m K)] |
| $k_m$ | Effective thermal conductivity [W/(m K)] |
| $K_p$ | Virus Partition coefficient |
| $K_n$ | Knudsen number of the particle |
| $L$ | Mask length [m] |
| $L_{ins}$ | Thickness of insulating material [m] |
| $\dot{m}$ | Mass flow rate of the fluid [kg/s] |
| $N_A$ | Avogadro's number |
| $P_m$ | Mask perimeter [m] |
| $P$ | Pressure [Pa] |
| $Pé$ | Péclet number |
| $\Delta P$ | Pressure drop across a single layer of mesh screen under oscillatory flow [Pa] |
| $q$ | Heat production per unit volume [W/m³] |
| $r$ | Virus thermal inactivation rate [mol/(L s)] |
| $R$ | Gas constant [J/(K mol)] |
| $Re_d$ | Reynolds number based on the wire diameter |
| $s_w$ | Fiber spacing [m] |
| $t$ | Time [s] |
| $T$ | Temperature [K] |
| $T_{amb}$ | Ambient temperature [K] |
| $U$ | Velocity [m/s] |
| $U_0$ | Superficial velocity of the fluid [m/s] |
| $U_{max}$ | Maximum air flow velocity within the mask [m/s] |
| $U_{vir}$ | Effective virus velocity [m/s] |
| $U_p$ | Particle velocity [m/s] |
| $V_{mask}$ | Mask volume [L] |
| $V_T$ | Tidal volume [L] |
| $\hat{V}$ | Reduced volume: ratio of mask volume to tidal volume |
| $W$ | Imposed electric power [W] |
| $x$ | Position along the mask [m] |
| $\alpha$ | Solidity of the screen |
| $\alpha'$ | Average solidity of the stacked fiber screens |
| $\eta_f$ | Single fiber efficiency |
| $\varphi$ | Porosity of the mesh |
| $\Lambda$ | Filtration efficiency of the porous mesh |
| $\lambda$ | Mean free path of a gas molecule in the fluid [m] |
| $\mu$ | Fluid dynamic viscosity [Pa s] |
| $\rho$ | Fluid density [kg/m$^3$] |
| $\tau$ | Period of human breath [s] |
| $exh$ | Exhalation |
| $f$ | Fluid phase (virus-laden air) |
| $inh$ | Inhalation |
| $s$ | Solid phase (copper mesh) |

## AUTHOR INFORMATION

### Corresponding Author

* Email: strano@mit.edu

## ACKNOWLEDGMENT

The authors acknowledge the Office of Naval Research (ONR) under award N00014-16-1-2144 for their support in the analysis of thermal materials design. S.F. and D. J. L. acknowledge support by the National Science Foundation Graduate Research Fellowship under Grant No. 1122374 and Grant No. 1745302, respectively.

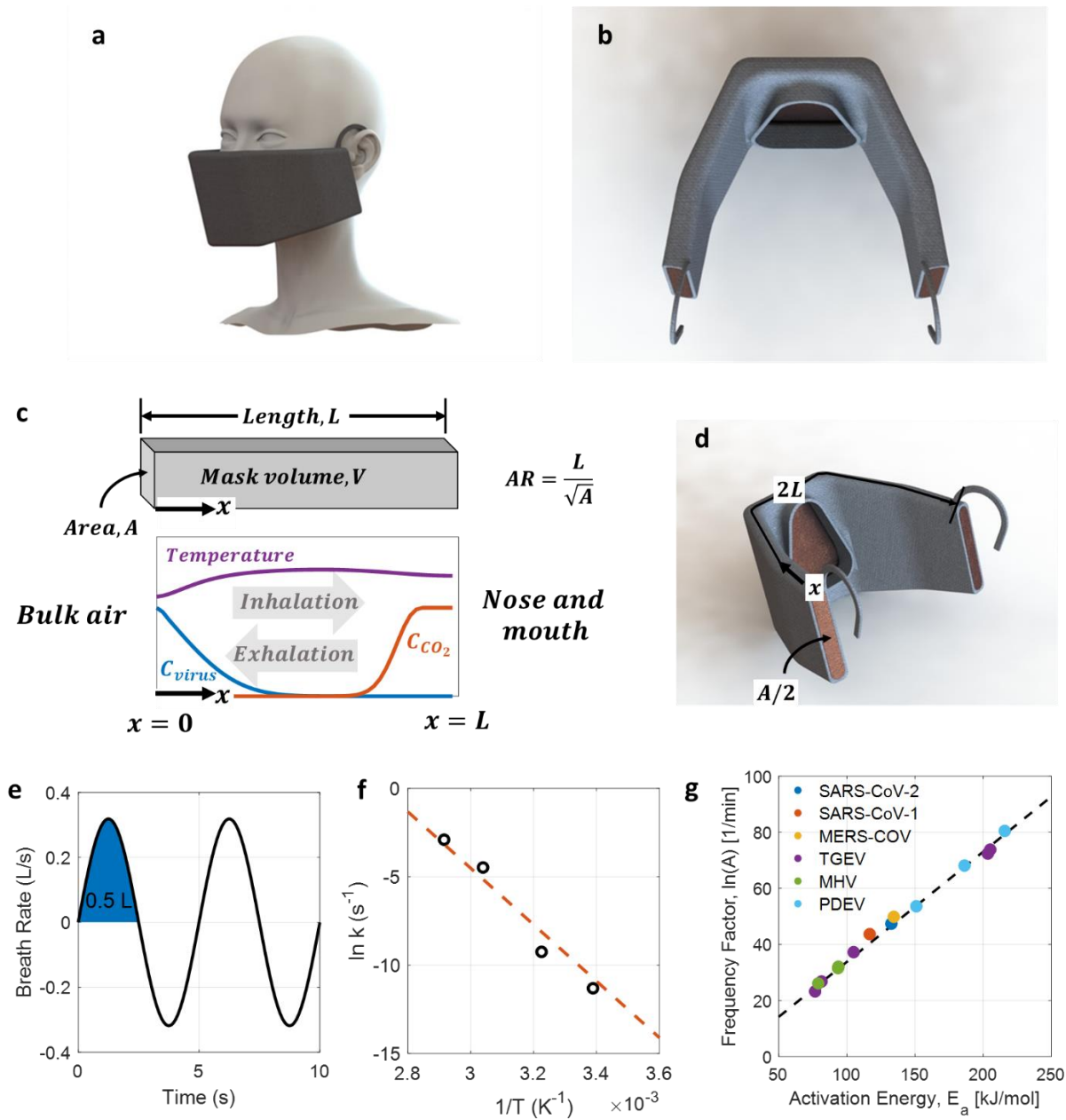

**Figure 1.** Heated mask design and problem formulation. (a) CAD drawing of a 0.3 L mask. The final proposed design is a mask with a volume of 0.3 L at an operating temperature of 90 °C, which can cause thermal inactivation of SARS-CoV-2 and operate comfortably with insulation and cooling. (b) A second CAD drawing of a 0.3 L mask, showing a porous 0.6 mm diameter copper mesh interior, 0.3 cm thick neoprene insulation, two inlets for air on the sides of the mask, and one

outlet to the nose and mouth. (c) The interior of the mask can be modelled as a 1D domain extending from bulk air at $x = 0$ to the nose and mouth at $x = x_{max}$. The mask length, cross-sectional area, and aspect ratio are defined. Temperature, virus concentration, and $CO_2$ concentration vary with position and time within the domain. Neumann and Danckwerts boundary conditions are enforced at $x = 0$ and $x = L$, and switch with inhalation and exhalation. (d) CAD drawing showing the 1D coordinate system and dimensions superimposed on the 3D mask.

(e) Sinusoidal approximation of human breath waveform, with a period of 5 seconds and a tidal volume of 0.5 L. (f) Thermal inactivation of SARS-CoV-2 follows first order kinetics, with a linear relationship between the natural log of the rate constant, ln k, and inverse temperature, 1/T. Data reproduced from Chin et al.[27] (g) Thermal inactivation of SARS-CoV-2 is well-characterized, with first-order rate parameters following a Meyer-Neldel rule in agreement with other coronaviruses like the first Severe Acute Respiratory Syndrome coronavirus (SARS-CoV-1), Middle East Respiratory Syndrome (MERS-CoV), Transmissible Gastroenteritis Virus (TGEV), Mouse Hepatitis Virus (MHV), and Porcine Epidemic Diarrhea Virus (PEDV). Data reproduced from Yap et al[28]

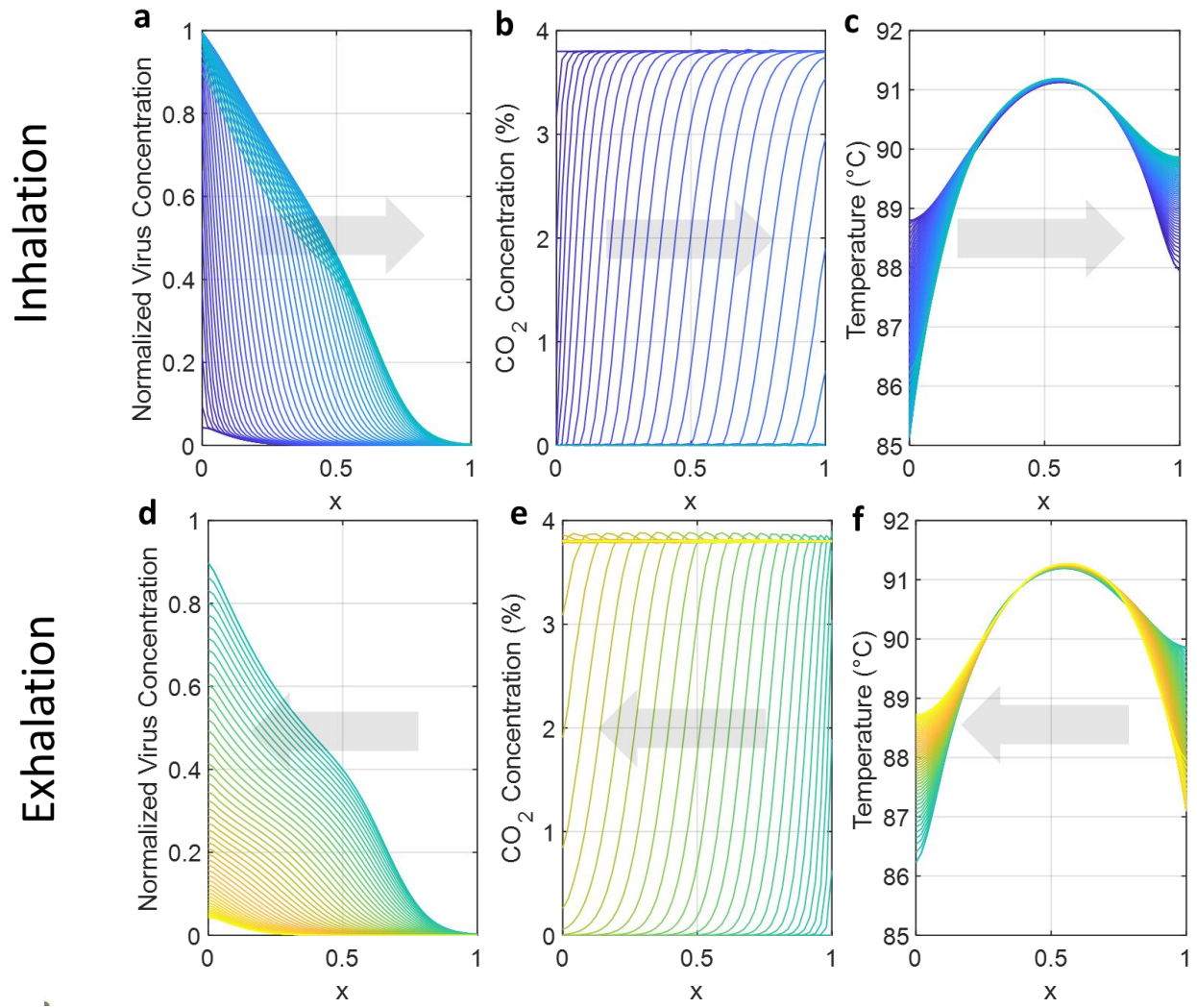

**Figure 2.** Viral concentration, $CO_2$ concentration, and temperature vary with position and time during a single 5 s cycle of inhalation (a,b,c) and exhalation (d,e,f) in the 1D mask model. The mask extends from bulk air ($x = 0$) to the mouth ($x = 1$). In this case, the mask volume is 0.3 L, the aspect ratio is 3, the partition coefficient ($K_p$) is 2, and the power is 18.89 W, generating a mean temperature of 90 °C. (a) During inhalation, virus is impeded and inactivated, and is generally not transported from the bulk to the mouth. (b) During inhalation, low-$CO_2$ concentration air is inhaled. (c) The mean temperature during inhalation is 90 °C, with slight variations. (d)

During exhalation, active virus in the mask is generally transported back to the bulk. (e) During exhalation, $CO_2$ is transported from the mouth through the mask. (f) The mean temperature during exhalation is 90 °C, with slight variations.

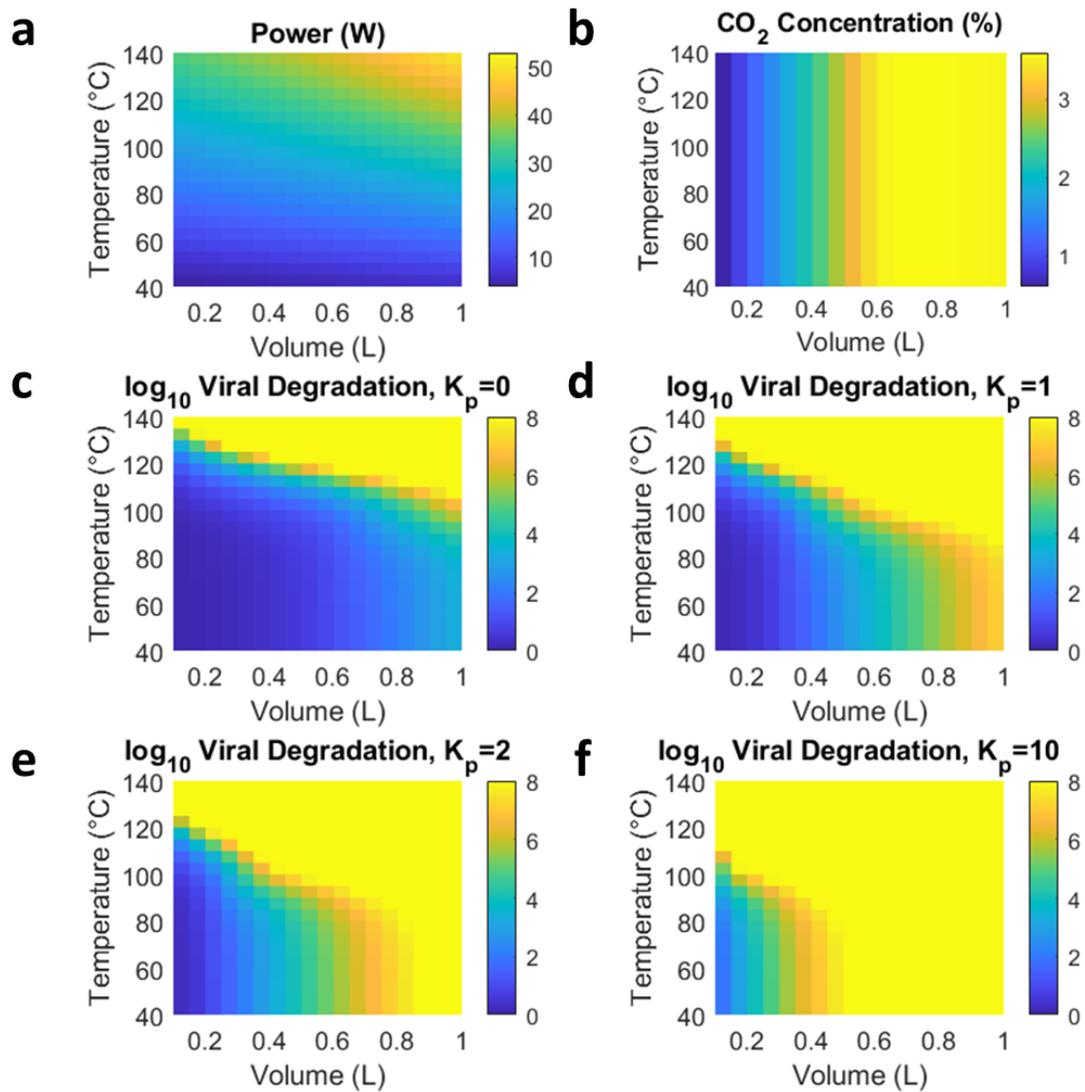

**Figure 3.** (a) Required power at steady state as a function of mask volume and operating temperature. Power requirements range from 3.79 W for a 0.1 L mask at 40 °C to 52.96 W for a 1 L mask at 140 °C. (b) Inhaled $CO_2$ concentration, averaged over a breath cycle. $CO_2$ concentration increases with increasing mask volume but is invariant with operating temperature. (c-f) $\log_{10}$ viral inactivation as a function of mask volume and mean operating temperature. Results are plotted for several values of the partition coefficient, $K_p$, which captures impedance of virus in the porous

mask mesh, with $K_p=0$ indicating that virus is transported with the same velocity as air in the mask, while $K_p \gg 1$ indicates substantial slowing of viral particles relative to air.

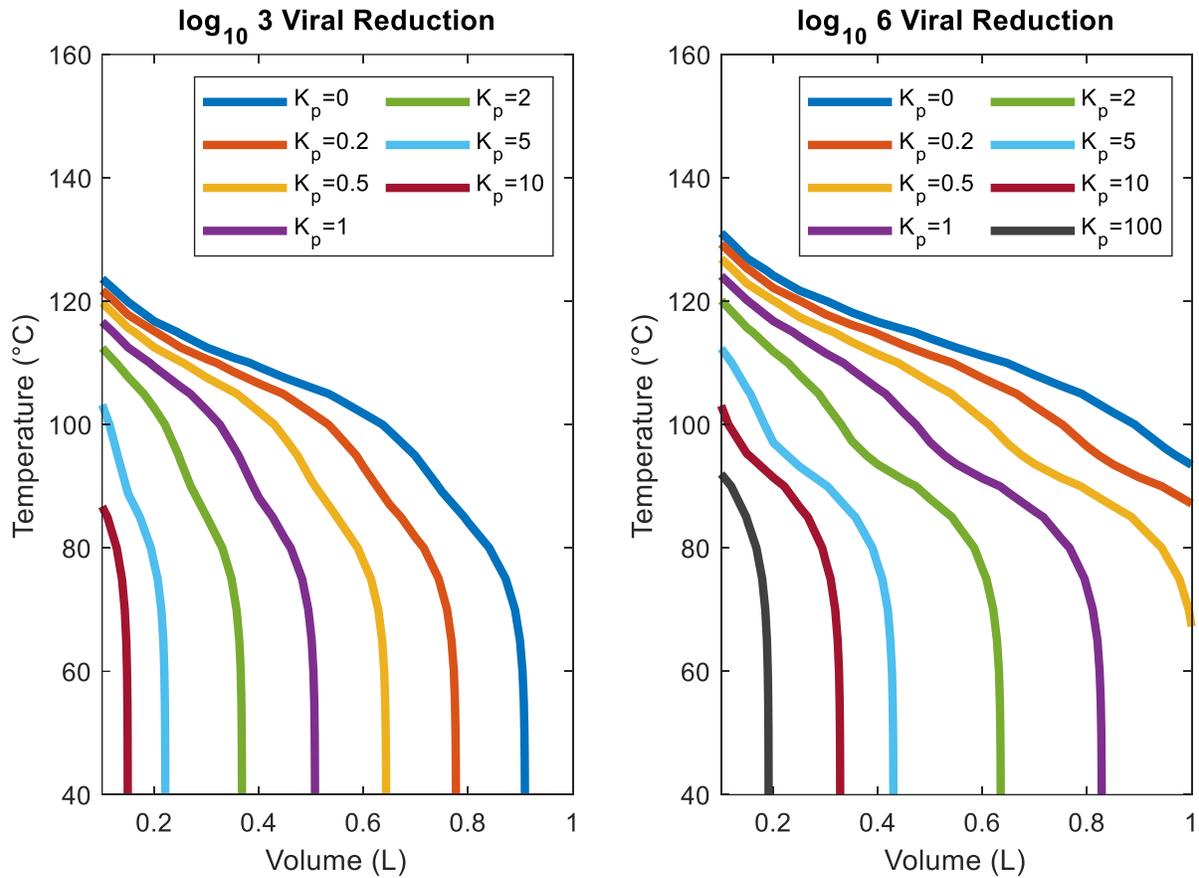

**Figure 4.** (a) Mask volume and operating temperature that are required to achieve 3-log reduction in SARS-CoV-2 concentration through a heated mask. Results are plotted as a function of a partition coefficient, $K_p$, which captures impedance of virus in the porous mask mesh, with $K_p = 0$ indicating that virus is transported with the same velocity as air in the mask, while $K_p \gg 1$ indicates substantial slowing of viral particles relative to air. A 3-log reduction in inhaled virus can be achieved in masks with volumes less than 0.4 L at temperatures below 100 °C if there is slight viral impedance in the mask ($K_p > 1$). (b) Mask volume and operating temperature that are required to achieve 6-log reduction, or sterilization, of SARS-CoV-2 concentration through a heated mask. A 6-log reduction in inhaled virus can be achieved in masks with volumes less than 0.4 L at temperatures below 100 °C with moderate viral impedance in the mask ($K_p > 2$).

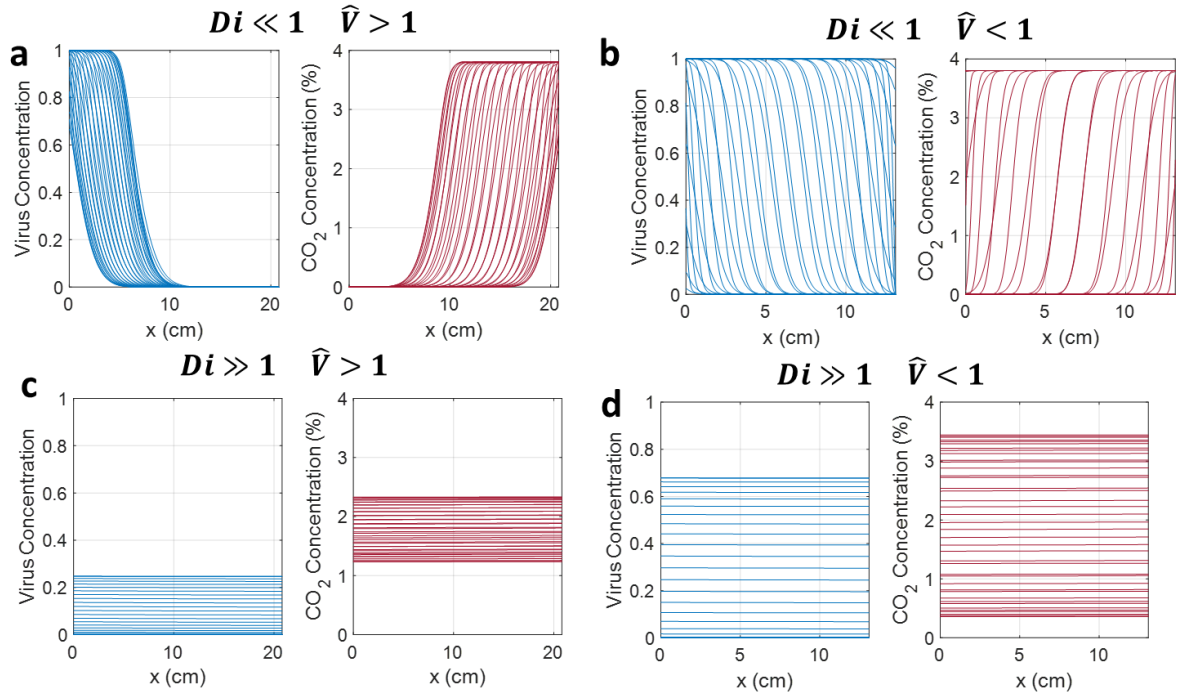

**Figure 5.** In analogy with reverse-flow chemical reactors, behavior of a heated mask can be split into different regimes depending on dispersion number Di and reduced volume $\hat{V}$. (a) The low Di, high $\hat{V}$ limit ("plastic bag limit") does not allow virus to reach the mouth but does not provide adequate ventilation. (b) The low Di, low $\hat{V}$ limit ("straw limit") provides ventilation but may allow viral inhalation depending on the viral inactivation rate. This is the most promising regime. (c) The large, well-mixed limit. Virus and $CO_2$ are well-mixed within the mask volume regardless of volume. (d) The small, well-mixed limit. At high Di, there is no bifurcation at $\hat{V} = 1$, so the small, well-mixed limit displays qualitatively similar behavior to the large, well-mixed limit, with temporal fluctuations in virus and $CO_2$ concentration during the breath cycle but near-zero spatial concentration gradients.

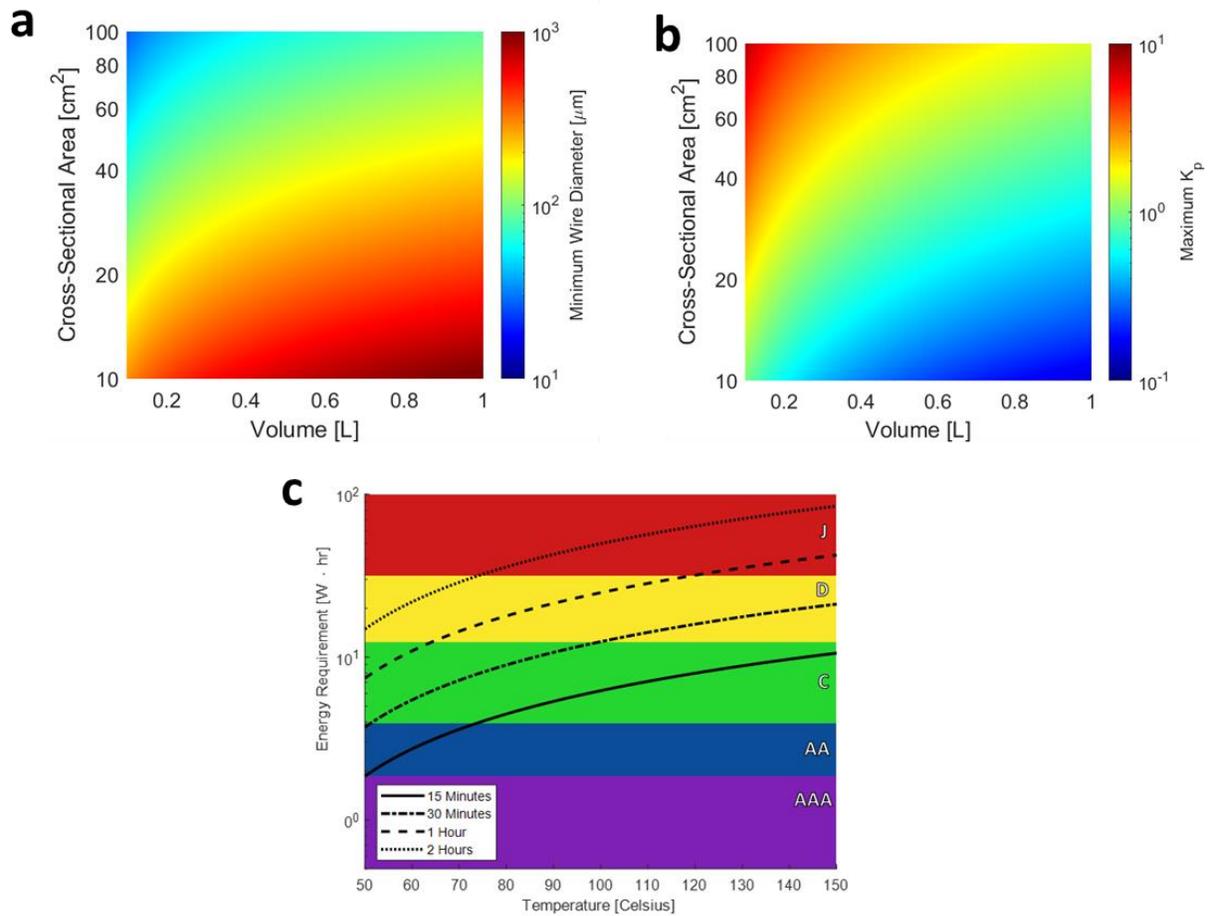

**Figure 6.** (a) Minimum allowable mesh diameter as a function of mask volume and cross-sectional area. The threshold mesh diameter induces a pressure drop of 60 Pa L$^{-1}$ s$^{-1}$ at the maximum breath flow rate of 0.4 liters per second. Mesh solidity is set to 0.1, with wire spacing equal to 3.2 times the wire diameter. Coarser meshes are required in larger masks to obey the pressure drop constraint, while meshes with higher cross-sectional area (and lower aspect ratio) can tolerate finer meshes without exceeding the maximum pressure drop. For the final design case, with a volume of 300 mL and a cross-sectional area of 50 cm$^2$, a mesh size of 0.1 mm is ideal. (b) Maximum achievable partition coefficient, $K_p$, as a function of mask volume and cross-sectional area. The partition coefficient is calculated as the ratio of the adsorption rate to the desorption rate, where $k_{ads}$ is calculated from fiber efficiencies while a conservative value of $k_{des} = 1\ s^{-1}$ is assumed. It is likely that a mask with a volume of 300 mL and an aspect ratio of 3 can achieve a $K_p$ of 2, allowing for 3-log reduction in SARS-CoV-2 transport. (c) Energy requirement of 300 mL mask versus average temperature. The energy requirement for a given amount of time is compared against the total power supply of standard commercial batteries that have the following power capacities and weights[54]: AAA, 1.87 W-hrs and 12 grams; AA, 3.9 W-hrs and 24 grams; C, 12.3 W-hrs and 65 grams; D, 31.5 W-hrs and 135 grams; and J, 540 W-hrs and 272 grams.

# Supplementary Information

## 1. Proposed Mask Design

Details for the proposed mask design are shown in **Table S1.**

**Table S1.** Proposed mask design

| Viral reduction | Log-3 (99.9%) |
|---|---|
| Volume (mL) | 300 mL |
| Temperature | 90 °C |
| $K_p$ | 2 |
| Wire diameter | 100 µm |
| Mesh solidity | 0.1 |
| Power | 20.7 W |
| Battery requirement | 4 9V rechargeable lithium-ion batteries |
| Battery lifetime | 1 hr |
| Mask mass | 350 g |
| Total mass (mask + battery) | 600 g |

## 2. Thermal Equilibrium Validation

The heat transfer model used in the simulation assumes that temperature difference between the solid part and the fluid is negligible. The assumption is validated by calculating the exact difference in temperature. Relevant dimensionless numbers are computed:

$$\Pr = \frac{c_p \mu}{k} = 0.7323 \qquad [1]$$

$\text{Re} = \frac{\rho_{air} V_{air} d_w}{\mu_{air}} = 0.843$  [2]

Heat transfer coefficient is approximated by empirical correlation for the average Nu number for forced convection over circular cylinder in cross flow.[1]

$\text{Nu} = \frac{hD}{k_f} = 0.989 Re^{0.330} Pr^{1/3}, 0.4 < R < 4$  [3]

Result yields that $h = Nu * \frac{k_f}{d_w} = 208 \frac{W}{m^2 K}$. For a 300 mL mask operating at a mean temperature of 90 °C, the averaged heat flux on the mesh surface and temperature difference between copper mesh and air are:

$q'' = \frac{P}{A_{mesh\ surface}} = 100.1 \frac{W}{m^2}$  [4]

$\Delta T = \frac{q''}{h} = 0.48\ K$  [5]

Since the temperature difference is within 0.7% of the temperature change within the mask, the thermal equilibrium approximation is identified as a valid assumption.

## 3. *Regenerative Heating Mask Design*

In the preliminary design, the mesh within the mask utilizes homogeneous material, which allows simple fabrication of the mask. However, due to the high thermal diffusivity of the mesh material, copper, the mask requires a high electric power input to maintain at desired temperature. In order to reduce the power requirement, a revised design is proposed: use extra "thermal masses" replaces the copper mesh on both ends of the heating channel. The thermal mass is able to capture more heat within the mass and reduce the heat transferred out of the mask. A low thermal diffusivity (low thermal conductivity, high heat capacity, $\alpha = \frac{k}{\rho c_p}$) in the thermal mass is desired. Low density and low cost are also favored features in our design.

Commercial nontoxic wire mesh materials, including aluminum, copper, yellow brass (70% Cu, 30% Zn), bronze (75% Cu, 25% Sn), and stainless steel (eg. T304), are considered as candidates for thermal masses. In addition, glass fiber is evaluated due to its low diffusivity, low density, and low cost. These materials and their relevant physical properties are shown in the table below. A fiberglass is a form of fiber-reinforced plastic where glass fiber is the reinforced plastic. The glass fiber is usually randomly arranged, flattened into a sheet, or woven into a fabric. Data presented is for E-Glass Fiber. Solid volume section presented in the table is obtained for 300 mL mask, with 1/4 of the mask occupied by thermal masses and solidity maintained at 0.1. In **Table S2.**, we identify that glass fiber has lowest thermal diffusivity and relatively low cost. The material is also lightweight.

**Table S2.** Material selection of extra thermal masses.

|  | k [W/mK] | $\rho$ [g/cm$^3$] | $c_p$ [J/gC] | $\rho c_p$ | $\alpha$ [mm$^2$/s] | weight [g] |
|---|---|---|---|---|---|---|
| Aluminum | 240 | 2.7 | 0.9 | 2.430 | 98.77 | 27 |
| Copper | 400 | 8.3 | 0.386 | 3.204 | 124.85 | 83 |
| Brass | 111 | 8.5 | 0.38 | 3.230 | 34.37 | 85 |
| Bronze | 26 | 7.2 | 0.435 | 3.132 | 8.30 | 72 |
| SS T304 | 16.2 | 8 | 0.5 | 4.000 | 4.05 | 80 |
| Glass fiber | 1.3 | 2.6 | 0.8 | 2.08 | 0.625 | 26 |

The reduction in power required was found by simulating the temperature distribution in the case of a 400 mL mask, with modified material properties and eliminated heat source within the volume of the mask replaced by the low thermal diffusivity material. The results indicated a significant reduce in required power as shown in **Figure S1**: with 1/8 of the total mesh replaced by glass fiber on each side of the channel, the electric power reduced from nearly 20.7 W to 15.8 W, while maintaining the mean temperature at 90°C.

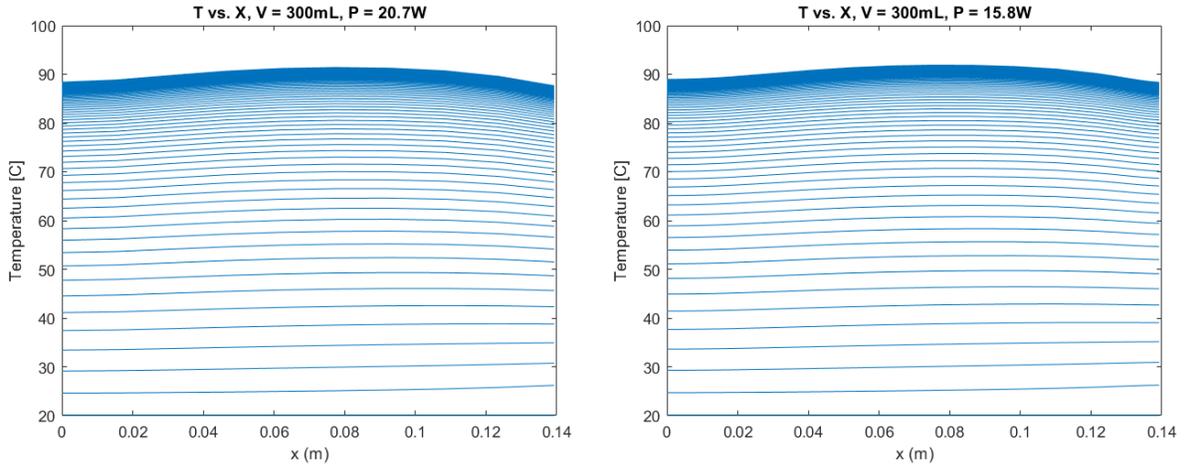

(a) No thermal mass: P = 20.7 W  (b) With fiberglass thermal mass: P = 15.8 W

**Figure S1**: Thermal Profile of Mask Containing Uniform Copper Mesh versus Low Thermal Diffusivity Fiberglass at the Inlet and Outlet.

### *4. 3D Mesh and Electrical Design*

The solidity in the simulation represents the 2D mesh design. It is thus important to show how the mesh is packed in 3D. The 3D mesh design is shown in **Figure S2**. The solidity of the 3D mesh is 0.13, which is considered close to the 2D mesh setting of 0.1 solidity. The electrical resistant is computed by Ohm's Law:

$$R = V^2/P \qquad [6]$$

$$P = \rho \frac{l}{A} \qquad [7]$$

where P is the applied electric power, V is the applied voltage, $\rho$ is the electrical resistivity of copper, $l$ is the length of the specimen, and A is the cross sectional area of the specimen. Current design of 9V rechargeable lithium-ion batteries has capacity of 600mAh and maximum discharge of 1C, i.e., maximum current of 600mA. If 4 batteries were connected in parallel, with copper mesh wire diameter of 0.1mm, the example design with an optimal mask volume of 300 mL at an operating temperature of 90 °C can be powered for 1 hour. A schematic of the mask mesh is shown below in Figure S2:

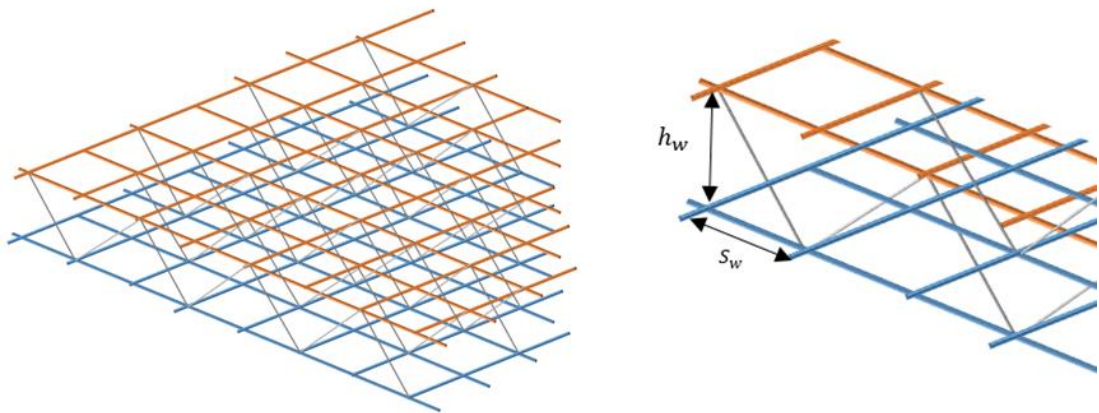

**Figure S2.** 3D mesh packing schematic. 2D mesh are stacked in layers and connected diagonally by zigzag inter-layer wires for supportive purpose. The height between each layer, $h_w$, is set to be the same as the wire spacing, $s_w$. The solidity of the 3D mesh is 86%, which is considered close to the 2D mesh.

**Supplemental Literature Cited**